\documentstyle[prl,aps,epsfig,twocolumn]{revtex}

\newcommand\be{\begin{eqnarray}}
\newcommand\ee{\end{eqnarray}}
\newcommand\nn{\nonumber}

\begin{document}
\title{Two-photon linewidth of light ``stopping'' via electromagnetically
induced transparency}
\author{Claudia Mewes and Michael Fleischhauer}
\address{Fachbereich Physik, Univ. Kaiserslautern, D-67663 Kaiserslautern,
Germany}
\date{\today}
\maketitle
\begin{abstract}
We analyze the two-photon linewidth of the recently proposed adiabatic
transfer technique for ``stopping'' of light
using electromagnetically
induced transparency (EIT). We show that a successful and
reliable transfer of excitation from light to atoms and back can be achieved
if the spectrum of the
input probe pulse lies within the initial transparency window of EIT, and
the two-photon detuning $\delta$ is less than the 
collective coupling strength (collective vacuum Rabi-frequency) 
$g\sqrt{N}$ divided
by $\sqrt{\gamma T}$, with $\gamma$ being the radiative decay rate, $N$ the 
effective number of atoms in the sample, and
$T$ the pulse duration. Hence in an optically thick medium 
light ``storage'' and retrieval is possible with high fidelity
even for systems with rather large two-photon detuning or
inhomogeneous broadening.
\end{abstract}

\pacs{42.50.-p,42.50.Gy,42.65.Tg,03.67.-a}


\section{Introduction}


One of the challenges of practical quantum information processing and
communication is the faithful storage and retrieval of an unknown quantum
state in a memory system \cite{deVinzenco}. Recently we have proposed 
a technique for 
a controlled transfer of the quantum state of a photon wavepacket to and from 
a collective atomic spin excitation \cite{Lukin00-ent,Fl00-pol,Fl02-pol}
using electromagnetically induced 
transparency (EIT) \cite{EIT} and Raman adiabatic passage
\cite{STIRAP}.  When a weak probe pulse and a
much stronger control field couple two metastable states of a 3-level atom
through a Raman transition in two-photon (but not necessarily in 
single-photon) resonance, the control field
renders an otherwise optically thick medium transparent. The induced
transparency is associated with a substantial reduction of the propagation
velocity of the probe pulse due to the formation of a coupled field-spin
excitation called dark-state polariton \cite{Fl00-pol,Fl02-pol}. 
Dynamically reducing the
intensity of the control field decelerates the polariton and 
can bring it to a full stop \cite{lui,phillips}.
When the velocity reaches zero,
the polariton is entirely
matter like and the quantum state of the original light pulse is completely
transferred to a collective spin excitation of the atomic ensemble. The
process is reversible and the quantum state can be transferred back to
a light pulse by re-accelerating the polariton, 
which can be an exact replica of the original one or
--- if desired --- can occupy different modes 
(different direction, carrier frequency
etc.) \cite{Juzeliunas,Scully}.

Essential for a high fidelity of the transfer process
is an explicitly time-dependent control field which 
varies in most parts adiabatically.
When the group velocity of the polariton approaches zero, so does the 
spectral width of transparency in EIT. 
Adiabatic following leads however to 
a narrowing of the spectral width of the probe pulse parallel to the
narrowing of the transparency window and thus there are no absorption 
losses during the slow-down provided 
the carrier frequency of the probe pulse and the control field are in
precise two-photon resonance \cite{Fl02-pol}. 
For a non-vanishing two-photon detuning the
pulse spectrum will move outside the transparency region at some finite
value of the group velocity. Thus the question arises what values
of the two-photon detuning, if any, are tolerable to maintain a 
sufficiently high fidelity of the quantum memory.
This question is of particular practical importance in gas experiments
with different pump and probe frequencies \cite{lui,phillips} 
or different propagation directions
of the fields since two-photon Doppler-shifts are then no longer negligible.
An estimate of the two-photon linewidth of light ``storage''
is furthermore interesting for 
applications in rare-earth doped solid-state materials with
inhomogeneously broadened two-photon transitions \cite{Hemmer02}.

In the present paper we analyze the two-photon linewidth of the
storage process based on an analytic perturbation theory and compare it
with exact numerical results. We will show that under otherwise favorable
conditions, the linewidth is given by the collectively 
enhanced coupling strength (vacuum Rabi-frequency) $g\sqrt{N}$, with
$N$ being the number of atoms divided by $\sqrt{\gamma T}$, with 
$\gamma$ being the excited-state decay rate and $T$ the characteristic pulse
duration. In an optically thick ensemble this
quantity can be large and thus rather large two-photon
detunings are tolerable.


\section{model}

\begin{figure}[ht]
\centerline{\epsfig{file=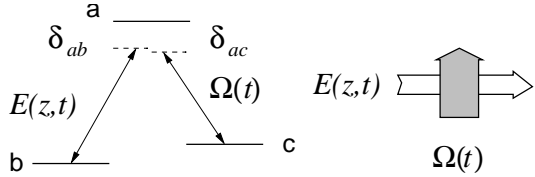,
width=7.0cm}}
\vspace*{2ex}
\caption{3-level $\Lambda$-type medium coupled to a control
field with Rabi-frequency $\Omega(t)$ and a probe field $E(z,t)$.}
\label{1}
\end{figure}

We consider the quasi 1-dimensional system
shown in Fig.~\ref{1}. A probe pulse with positive frequency part of 
the electric field $E^{(+)}$
couples the transition between the ground state $|b\rangle$ and
the excited state $|a\rangle$. $\delta_{ab} =\omega_{ab}-\nu$ 
is the detuning between the carrier frequency $\nu$ and the atomic 
transition frequency $\omega_{ab}$. 
The upper level $|a\rangle$ is coupled to the stable state 
$|c\rangle$ via a coherent control field
with Rabi-frequency $\Omega$. $\delta_{ac}=\omega_{ac}-\nu_c$ is the 
corresponding detuning of the coupling transition. 
The Rabi-frequency of the coupling field is assumed to be large compared to
that of the probe pulse and undepleted. Furthermore we assume that $\Omega$
is only a function of time. This can be realized either by
perpendicular incidence of the control field or, in the case of
co-propagating fields, if the group velocity of the probe pulse is at all
times much less than that of the coupling field. In the latter case 
retardation effects of the control field can be disregarded.


Within the rotating wave approximation 
the interaction between the atoms and the fields can be described by
density matrix equations in a
rotating frame
\begin{eqnarray}
\dot{\rho}_{aa} &=&-\gamma_a {\rho}_{aa}
-ig\Bigl({\cal E}^*{\rho}_{ab}-c.c.\Bigr)\nn\\
&&-i\Bigl(\Omega^*{\rm e}^{-i k z}{\rho}_{ca}-c.c.\Bigr),\label{rho_aa}\\
\dot{\rho}_{bb} &=&\gamma_{a\to b} {\rho}_{aa}
+ig\Bigl({\cal E}^*{\rho}_{ab}-c.c.\Bigr),\\
\dot{\rho}_{cc} &=&\gamma_{a\to c} {\rho}_{aa}
+i\Bigl(\Omega^*{\rm e}^{-i k z}{\rho}_{ac}-c.c.\Bigr),\\
\dot{\rho}_{ab} &=&-\Bigl(\gamma_{ab}+i\delta_{ab}\Bigr) {\rho}_{ab}
+ig{\cal E}\Bigl({\rho}_{bb}-{\rho}_{aa}\Bigr)\nn\\
&&+i\Omega{\rm e}^{i k z}{\rho}_{cb},\\
\dot{\rho}_{ac} &=&-\Bigl(\gamma_{ac}+i\delta_{ac}\Bigr) {\rho}_{ac}
+i\Omega{\rm e}^{i k z}
\Bigl({\rho}_{cc}-{\rho}_{aa}\Bigr)\nn\\
&&
+ig{\cal E}{\rho}_{bc},\\
\dot{\rho}_{cb} &=&
-i\Bigl(\delta_{ab}-\delta_{ac}\Bigr)\rho_{cb}
+i\Omega^*{\rm e}^{-i\Delta k z}{\rho}_{ab}
-ig {\cal E}{\rho}_{ca},\label{rho_cb}
\end{eqnarray}
where 
$\gamma_a=\gamma_{a\to b}+\gamma_{a\to c}$ and
$\gamma_{a\to b}, \gamma_{a\to c}$ denote longitudinal
and 
$\gamma_{\mu\nu}$ transverse decay rates, and 
$k=\nu/c$ is the wavenumber of the probe field propagating in the
$z$ direction.
We have also introduced the dimensionless slowly
varying field amplitude
\be
E^{(+)}(z,t) = \sqrt{\frac{\hbar \nu }{2\varepsilon_0 V}}
\, {\cal E}(z,t)\, {\rm e}^{i\frac{\nu}{c}(z-ct)},
\ee
with $V$ being the interaction volume. 
To simplify the following analytic
discussion it is useful to consider a resonant probe
field, i.e. to assume $\delta_{ab}=0$, while keeping $\delta_{ac}\ne 0$.

The evolution of the probe pulse  can be described in slowly varying 
amplitude approximation (SVEA) by the propagation equation
\begin{eqnarray}
\left(\frac{\partial}{\partial t}+c\frac{\partial}{\partial z}\right)
{\cal E}(z,t)= { i} g N\, \rho_{ab}(z,t).\label{MWE}
\end{eqnarray}
Here $g=\wp\sqrt{\frac{\nu}{2\hbar\epsilon_0 V}}$ is the atom-field 
coupling constant proportional to the dipole moment $\wp$ of the probe
transition. 

It was shown in Ref.\cite{Fl00-pol,Fl02-pol} 
that the propagation of elm. pulses 
in an EIT medium can be be most easily understood in terms of 
polariton-like fields, which are superpositions of the (dimensionless)
electric field amplitude ${\cal E}$ and the spin coherence $\rho_{cb}$ 
between the two lower levels. There are two types of polaritons called
dark-state ($\Psi$) and bright-state ($\Phi$) polaritons:
\be
\Psi(z,t) &=& \cos\theta(t) {\cal E}(z,t) -\sin\theta(t)\sqrt{N}\rho_{cb}
\, {\rm e}^{i k z},\label{Psi-eq}\\
\Phi(z,t) &=& \sin\theta(t) {\cal E}(z,t) +\cos\theta(t)\sqrt{N}\rho_{cb}
\, {\rm e}^{i k z},\label{Phi-eq}
\ee
The mixing angle $\theta$ in eq.(\ref{Psi-eq})
 and (\ref{Phi-eq}) is defined via
$\tan\theta(t)=g\sqrt{N}/\Omega(t)$, where we have assumed 
without loss of generality a real Rabi-frequency of the drive field 
$\Omega=\Omega^*$.

One can transform the equations of motion for the electric field and
the atomic variables into the polariton variables. In the 
approximation of low probe field intensities --- which is most interesting for
quantum memory purposes --- one finds
\be
&&\biggl[\frac{\partial}{\partial t} +c\cos^2\theta
\frac{\partial}{\partial z}-i\, \delta\sin^2\theta\biggr]\,
\Psi =\nonumber\\
&&\qquad-\left[\dot\theta+\sin\theta
\cos\theta\, \left(c\frac{\partial}{\partial z}+i\, \delta\right)\right]
\Phi\label{Psi-full},
\ee
and 
\be 
&&\Phi =\frac{\sin\theta}{g^2 N}\times \label{Phi-full}\\
&&\quad\times
\biggl(\frac{\partial}{\partial t}+\gamma\Bigr)\biggl[\tan\theta 
\left(\frac{\partial}{\partial t}-i\delta\right)\biggr]
\Bigl(\sin\theta\,\Psi-\cos\theta\,\Phi\Bigr),\nonumber
\ee
with $\gamma=\gamma_{ab}$. One here 
has to keep in mind that the mixing angle $\theta$ 
is a function of time.


\section{polariton dynamics for finite two-photon detuning}


In order to find approximate analytic solutions of these equations
we introduce a characteristic time $T$ of the transfer and a corresponding
dimensionless time, length and detuning. We can then identify two
expansion parameter, $\epsilon_1=(g\sqrt{N} T)^{-1}$,
defining an adiabatic expansion, and $\epsilon_2=\delta/(g\sqrt{N})$
characterizing the magnitude of the two-photon detuning.
Furthermore we assume for simplicity
that $T$ is much larger than the decay time,
i.e. $\gamma T \gg 1$, which is usually well satisfied for
conditions of light ``stopping''.

In the adiabatic limit, i.e. up to first order of $\epsilon_1$, and
for small detuning, i.e. in first order of $\epsilon_2$ one finds:
\be
\biggl(\frac{\partial}{\partial t}+ c\cos^2\theta
\frac{\partial}{\partial z}-i\delta\sin^2\theta(t)\biggr)\,
\Psi^{(0)}(z,t) &=&0  \label{Psi-0},\\
\Phi^{(0)}(z,t) &=&0 .\label{Phi-0}
\ee
In this limit the bright-state polariton is not exited.
The dark-state polariton propagates with a form-stable
envelope and with instantaneous velocity $v_{\rm gr}(t)=c\cos^2\theta(t)$.
The small two-photon detuning simply causes a time dependent phase shift
(chirp) of the pulse:
\be
\Psi^{(0)}(z,t) &=&\Psi\Bigl
(z-c\int_0^t\!\! {\rm d}t^\prime \cos^2\theta(t^\prime),0\Bigr)\nonumber\\
&&\times
\exp\left\{i\delta \int_0^t\!\! {\rm d}t^\prime \sin^2\theta(t^\prime)
\right\}
\ee

\begin{figure}[ht]
\centerline{\epsfig{file=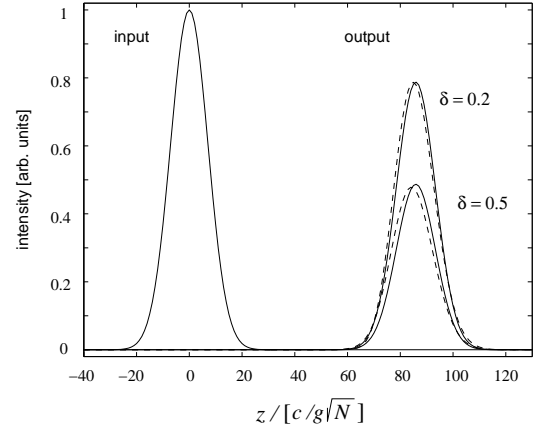,
width=7.0cm}}
\vspace*{2ex}
\caption{Polariton intensity at  $t=0$ (time in units of $(g\sqrt{N})^{-1}$)
(full line at left side) and after
storage and release ($t=150$) for finite two-photon detuning
$\delta=0.2 g\sqrt{N}$ and $0.5 g\sqrt{N}$, and for $\gamma/g\sqrt{N}=1$. 
Shown are the analytic expression
(full line) and a numerical result obtained by solving the 
full Maxwell-Bloch equations in SVEA (dashed line).
The mixing angle $\theta(t)$ was rotated according to
$\cot\theta(t) = 100-50 \tanh[0.1 (t -15)]+50 \tanh[0.1 (t -125)]$}
\label{dynamics}
\end{figure}

In next order of perturbation in $\epsilon_1$ and $\epsilon_2$
a longer calculation gives a finite excitation of the bright-state
polariton due to non-adiabatic couplings and the non-vanishing
 two-photon detuning
\be
\Phi^{(2)} &=& \frac{\gamma\sin^2\theta}{g^2 N}\Biggl[\dot\theta
-i \delta \sin\theta\cos\theta\Biggr]\Psi^{(2)}\nn \\
&& -\frac{\gamma \sin^3\theta\cos\theta}{g^2 N}
c\frac{\partial}{\partial z}\Psi^{(2)}.\nn
\ee
Here we made use of the fact that in the same order of approximation 
$\partial\Psi/\partial t$ can be replaced by 
$i\delta\sin^2\theta\Psi-c^2\cos^2\theta(\partial\Psi/\partial z)$.
Substituting this result into the right hand side of eq.(\ref{Psi-full})
gives the equation of motion
\be
&&\biggl(\frac{\partial}{\partial t}+ c\cos^2\theta
\frac{\partial}{\partial z}-i\delta\sin^2\theta(t)\biggr)
\Psi^{(2)} = \nn\\
&&\quad -\biggl(A_0(t) +\delta^2 A_1(t)\biggr) \Psi^{(2)}
-i\delta B_0(t) c\frac{\partial}{\partial z}\Psi^{(2)}\label{Psi-2}\\
&&\quad -C_0(t)\, c^2\frac{\partial^2}{\partial z^2}
\Psi^{(2)}\nn
\ee
with
\be
A_0(t) &=& \frac{\gamma}{g^2 N} \dot\theta^2\sin^2\theta,\\
A_1(t) &=& \frac{\gamma}{g^2 N}  \sin^4\theta\cos^2\theta,\\
B_0(t) &=& -\frac{2\gamma}{g^2 N}\sin^4\theta\cos^2\theta,\\
C_0(t) &=& -\frac{\gamma}{g^2 N}\sin^4\theta\cos^2\theta ,
\ee
Since all coefficients $A_0,\dots, C_0$ depend only on time,
eq.(\ref{Psi-2}) can easily be integrated by Fourier-transformation
in space. To illustrate the accuracy of the approximations
we have compared in Fig.\ref{dynamics} the analytical expression
for the polariton intensity obtained from (\ref{Psi-2})
after storage and release with an exact numerical result. One
recognizes rather good agreement even for values of
$\epsilon_2$ as large as $0.5$.

$A_0$ is a dissipative loss
term due to non-adiabatic couplings which restricts the
speed of rotation between field-like and matter-like
behavior of the polariton \cite{Fl02-pol}.
$C_0$ causes dissipative losses of the high-frequency
components of the polariton.  It restricts the spectral
width of the light pulse to within the initial transparency
window of EIT (see \cite{Fl02-pol} for details).


\begin{figure}[ht]
\centerline{\epsfig{file=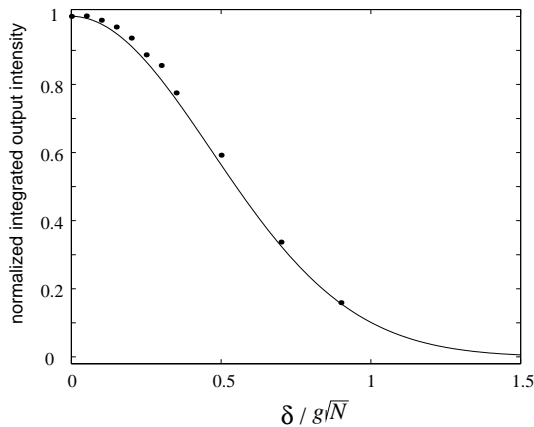,
width=7.0cm}}
\vspace*{2ex}
\caption{Integrated output intensity of the polariton
as function of $\delta/(g\sqrt{N})$ normalized to its value
at $\delta=0$ obtained from a numerical solution of the Maxwell-Bloch
equations (large dots) 
for the Gaussian input pulse and $\theta(t)$ of Fig.\ref{dynamics}
and for $\gamma/(g\sqrt{N})=1$.
The full line shows the analytic approximation (\ref{pol-loss}).
}
\label{detuning}
\end{figure}

The terms most essential for the two-photon linewidth of 
the ``light storage'' are $A_1$ and $B_0$. $A_1$ is a
frequency-independent dissipation term and $B_0$ 
accounts for deviations from this value depending on the
$k$-space Fourier-frequency of the polariton.
The frequency-independent losses due to a finite two-photon
detuning are thus given by
\be
\exp\left\{-\frac{\gamma\delta^2}{g^2 N}\int_0^\infty
\!\!{\rm d}t\, \cos^2\theta(t)\sin^4\theta(t)\right\}.\label{pol-loss}
\ee
There are contributions to the losses only for times when
neither $\sin\theta$ nor $\cos\theta$ is zero. $\sin\theta=0$
corresponds to the limit of an infinite drive-field Rabi-frequency.
In this case the Autler-Townes splitting of the excited state
by the drive field suppresses any absorption despite the finite two-photon
detuning. If, on the other hand, $\cos\theta=0$, the polariton 
is entirely matter-like
and thus a two-photon detuning is of no relevance. If we denote
the characteristic time for rotating $\theta$ from $0$ to $\pi/2$,
i.e. the time of transfer of the polariton from a pure electromagnetic
to a pure spin excitation, by $T$, i.e.
\be 
T\equiv \int_0^\infty
\!\!{\rm d}t\, \cos^2\theta(t)\sin^4\theta(t),
\ee
we find the following
condition for the two-photon detuning
\be
\delta \ll \delta_{\rm 2ph} = \frac{g\sqrt{N}}{\sqrt{\gamma T}}
\ee
One recognizes that the two-photon linewidth of the ``light storage''
process $\delta_{\rm 2ph}$ is proportional to the collective
Rabi-frequency. Thus in an optically thick medium rather large
two-photon detunings can be tolerated. This is illustrated in 
Fig.\ref{detuning}, where we have plotted the normalized integrated
intensity of the polariton after storage and release, obtained
from a numerical solution of the SVEA Maxwell-Bloch equations
(\ref{rho_aa}-\ref{rho_cb}) and (\ref{MWE}), as function of 
$\epsilon_2=\delta/g\sqrt{N}$.


\section{summary}


We have derived an 
analytic expression for the two-photon linewidth of
storage and retrieval of light pulses by
stimulated Raman adiabatic passage 
which is in excellent agreement with exact numerical simulations.
We have shown that the transfer process 
tolerates a rather large two-photon
detuning if the medium is optically thick. This allows the application
of the light-storage technique to systems with a non-vanishing inhomogeneously
broadened two-photon transition.


\section*{Acknowledgement}


This work was supported by the Deutsche Forschungsgemeinschaft
under grant Fl210/10 within the program on quantum information. 
C.M. acknowledges support by the Studienstiftung des deutschen Volkes.


\def\etal{\textit{et al.}}

\end{document}